\documentclass[floatfix, amsmath,reprint, amssymb,aps,prb,twocolumn, superscriptaddress]{revtex4-2}
\usepackage{mathtools,amsmath,graphicx,units}
\usepackage{dcolumn}
\usepackage{bm}
\usepackage{xcolor}
\usepackage{physics}
\usepackage{float}
\usepackage{epstopdf}
\usepackage{enumitem}
\usepackage[normalem]{ulem}
\usepackage[plainpages=false,pdfpagelabels,colorlinks=true,linkcolor=red,urlcolor=blue,citecolor=blue,pdftitle={Title},pdfauthor={},pdfdisplaydoctitle=true,pdfduplex=DuplexFlipLongEdge]{hyperref}
\usepackage{cleveref}
\usepackage[T1]{fontenc} 
\usepackage{orcidlink}

\newcommand{\revise}[1]{\textcolor{blue}{#1}}

\usepackage[makeroom]{cancel}

\begin{document}

\title{
Lyapunov spectrum scaling transition for quasiperiodic nonlinear unitaries 
}
\author{Xiaodong Zhang,\orcidlink{0000-0002-9173-7109}}
\email{xiaodongzhang2021@gmail.com}
\affiliation{Center for Theoretical Physics of Complex Systems, Institute for Basic Science, Daejeon 34126, Republic of Korea}
\affiliation{
Lanzhou Center for Theoretical Physics, Key Laboratory of Theoretical Physics of Gansu Province, \\
and Key Laboratory of Quantum Theory and Applications of MoE, Lanzhou University, Lanzhou, Gansu 730000, China
}
\affiliation{Shandong Key Laboratory of Space Environment and Exploration Technology, College of Physics and Electronic Engineering, Qilu Normal University, Jinan 250200, China}
\author{Barbara Dietz,\orcidlink{0000-0002-8251-6531}}
\email{bdietzp@gmail.com}
\affiliation{Center for Theoretical Physics of Complex Systems, Institute for Basic Science, Daejeon 34126, Republic of Korea}
\affiliation{Basic Science Program, Korea University of Science and Technology (UST), Daejeon 34113, Republic of Korea}
\affiliation{Max Planck Institute for the Physics of Complex Systems, N\"othnitzer Str. 38, 01187 Dresden, Germany}
\affiliation{TU Dresden, Institute of Theoretical Physics, 01062 Dresden, Germany}
\author{Sergej Flach}
\email{sergejflach@googlemail.com}
\affiliation{Center for Theoretical Physics of Complex Systems, Institute for Basic Science, Daejeon 34126, Republic of Korea}
\affiliation{Center for Trapped Ions Quantum Science, Institute for Basic Science, Daejeon 34126, Republic of Korea}
\affiliation{Basic Science Program, Korea University of Science and Technology (UST), Daejeon 34113, Republic of Korea}
\affiliation{Centre for Theoretical Chemistry and Physics, The New Zealand Institute for Advanced
Study (NZIAS), Massey University Albany, Auckland 0745,
New Zealand}

\date{\today}
\begin{abstract}
We study the Lyapunov spectrum scaling of thermal weakly-nonlinear unitary maps in the presence of quasiperiodic potentials. We search for the crossover from long-range to short-range scaling as the localization length $\xi$ decreases and compare the details to the case of uncorrelated Anderson disorder [Phys. Rev. Res. {\bf 6} L012064 (2024)].
A comparative statistical analysis of the eigenstates for the linear case shows that quasiperiodicity has a stronger localization impact at the same value of $\xi$. Therefore we expect that the scaling crossover should be enhanced as well. However, the numerical analysis shows that it is strongly delayed as compared to Anderson disorder, and is observed at anomalously small values of $\xi$. 
These findings hint at the potential impact of long range correlations of quasiperiodic localized eigenstates, which persist in the presence of interactions even in the case of integrability breaking and thermalization. 
\end{abstract}    
\maketitle
\section{Introduction}
Thermalization of nonintegrable many-body systems comes with characteristic timescales that diverge in proximity to integrable limits\cite{book:huang87,bel_weak_2005,rigol_breakdown_2009,DanieliCampbellFlachPRE2017,danieli_dynamical_2019,mithun_dynamical_2019,mithun_fragile_2021}. 
Ergodicity relates time and phase space averages of observables, which is the prime objective of any experiment. To quantify associated slowing-down processes, a possible attempt would be to study finite time averages of observables. This approach suffers from the ambiguity in the choice of an observable and the resulting plethora of time scales. With some care \cite{malishava_lyapunov_2022,malishava_thermalization_2022,danieli_dynamical_2019} one may ensure that proper ergodization timescales are identified
\cite{goldfriend_equilibration_2019,ganapa_thermalization_2020,malishava_thermalization_2022, baldovin_statistical_2021}. And yet the outcome is leaving us substantial lack of satisfaction, even though it might be of use for the development of microscopic theories.

A second approach is to employ the fact that nonintegrable classical many-body dynamics will be in general chaotic. The computation of a Lyapunov spectrum and the study of its scaling is independent of any observable and basis choice. Thus an unambiguous and universal way to address the above question is to simply study its scaling behavior upon approaching integrable limits.

Let us assume that the integrable limit possesses a unique choice of action and angle coordinates. This is indeed typically the case for most known integrable models. Here, we explicitly exclude degenerate ones from our considerations since they have a continuum of action angle choices (also called 'superintegrable' models).  Writing the weak nonintegrable
perturbation in the unique action angle basis and analyzing the resulting network spanned by the nonintegrable perturbation among the actions of the integrable limit, revealed that typical physics-related models result in two distinctively different networks, namely long-range networks (LRN) or short-range networks (SRN) \cite{danieli_dynamical_2019,mithun_dynamical_2019,mithun_fragile_2021}. The former ones are obtained for translationally invariant lattices with weak nonlinearities constituting mean field analogues of weak two-body interactions since the linear limit is integrable and the normal modes, which result in conserved actions, are extended over the entire system volume, and therefore connected in a long-range fashion once nonlinearity is added. The latter ones can be easily obtained at finite strength of nonlinearity by tuning the nearest-neighbour coupling strength on the lattice down to zero, resulting in a limiting integrable case of uncoupled oscillators, rotors or spins etc.

The subsequent study of Lyapunov-spectrum scaling (see Ref. \cite{constantoudis_nonlinear_1997} for early attempts) revealed that the spectra scale in qualitatively different ways for both network types. Surely the largest Lyapunov exponent always scales to zero upon approaching integrable limits in any of the networks. Yet, when rescaling the Lyapunov spectrum in units of the largest one it becomes (a) stationary for LRN cases, and (b) exponentially vanishing for SRN cases \cite{malishava_lyapunov_2022,malishava_thermalization_2022,lando2023thermalization,zhang2025observation,patra2026trotter,patra2026thermalization}.

A lattice with disorder will break translational invariance. At the linear limit it will be still integrable, but may, depending on the dimension, have all normal modes being Anderson localized \cite{anderson_absence_1958}. Adding nonlinearity will now couple one of these modes only with a finite number of other modes, which depends on the strength of disorder and localization. It is therefore expected, and has been observed~\cite{zhang2024thermalization} that weakly nonlinear lattice models will show a transition from LRN scaling to SRN scaling upon adding and increasing the strength of disorder. This transition depends sensitively on the finite system size as in any computational study and the comparative value of the localization length.

In this work we replace random disorder by a quasiperiodic potential which is a broadly accepted pathway to experimentally study localization physics e.g. with cold atoms and light \cite{lucioni2011observation,doi:10.1126/science.aaa7432,PhysRevLett.103.013901}. We expect that the removal of eigenvector fluctuations which are typical for disordered cases may result in a clearer transition from the LRN to SRN. We first introduce the model details, proceeding to the comparative analysis of eigenvectors of the linear case for both disorder and quasiperiodic cases. We then introduce the computation of Lyapunov spectra and proceed with the comparative study of the scaling in the disorder and quasiperiodic cases.
 
\section{Model}
\begin{figure}[]
    \includegraphics[width=0.45\textwidth]{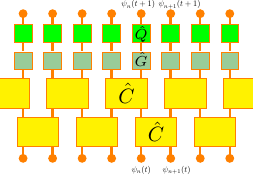}
    \caption{ A schematic representation of the unitary circuit map. The evolution proceeds from bottom to top and comprises alternating large yellow blocks, small dark green blocks, and light green blocks. These represent unitary matrices \(\hat{C}\), which are parameterized by the angle \(\theta\), local nonlinear transformation maps \(\hat{G}\) (dark green), governed by the nonlinearity strength \(g\), and local quasiperiodic potential maps \(\hat{Q}\) (light green), characterized by site-dependent phases \(\epsilon_n\). The states \(\psi_n(t)\) and \(\psi_{n+1}(t)\) evolve to \(\psi_n(t+1)\) and \(\psi_{n+1}(t+1)\), respectively, through successive applications of these transformations. Each complete time step consists of four substeps.}
    \label{map}
\end{figure} 
We employ a nonlinear unitary circuit map \cite{malishava_lyapunov_2022,malishava_thermalization_2022,zhang2024thermalization,zhang2025observation} with an additional quasiperiodic potential \cite{mallick2023intermediate}. 
Unitary circuit maps belong to the class of quantum random walks which are defined on specific networks or graphs, and  were introduced by Aharonov
et al. \cite{PhysRevA.48.1687}. They are used 
to implement a variety of quantum logical elements and
protocols \cite{kempe2003quantum,Venegas-Andraca2012}.  Unitary maps can be viewed as realizations of Floquet lattice models, and were implemented in numerous experimental setups,
including quantum optical systems \cite{peruzzo2010quantum}, ion traps \cite{schmitz2009quantum}, and
nuclear magnetic resonance systems \cite{du2003experimental}. 

\revise{Our} model is defined on a one-dimensional lattice consisting of \( N \) unit cells. Each of them comprises two sites. These are labeled by indices \( n \in\{ 1,2,\ldots,2N\} \), and
each site \( n \) is represented by a complex component \( \psi_n \). The initial state of the lattice is described by the vector \(\boldsymbol{\psi} = (\psi_1, \psi_2, \dots, \psi_{2N})\). The system evolves in a phase space of dimension \( 4N \), as each complex component contributes two real variables. We impose periodic boundary conditions \( \psi_{2N+1} = \psi_1 \). The evolution of a deterministic trajectory defined by its initial values and generated by iterative applications of the unitary map
\begin{equation}
\hat{U}_n = \hat{Q}_n \hat{G}_n \hat{C}_{n-1,n} \hat{C}_{n,n+1},  \quad  n\, {\rm odd}
\end{equation}
where each iteration corresponds to one time step. The unitary map \( \hat{U}_n \) is constructed from four key unitary components, namely two successive rotation operations \( \hat{C} \), a nonlinear operation \( \hat{G} \), and a quasiperiodic potential \( \hat{Q} \) as illustrated in Fig.~\ref{map}. The rotations \( \hat{C} \) are unitary transformations that act on pairs of neighboring sites \( n \) and \( n+1 \), defined as:
\begin{equation}
\hat{C}_{n,n+1} \begin{pmatrix} \psi_n \\ \psi_{n+1} \end{pmatrix} = 
\begin{pmatrix} 
\cos\theta & \sin\theta \\ 
-\sin\theta & \cos\theta 
\end{pmatrix} \begin{pmatrix} \psi_n \\ \psi_{n+1} \end{pmatrix},\, \theta\in [0,\pi/2].
\end{equation}

The nonlinear operation \( \hat{G}_n \) introduces site-dependent phase shifts proportional to the squared norm \( |\psi_n|^2 \) on site $n$, 
\begin{equation}
\hat{G}_n \psi_n = e^{i g |\psi_n|^2} \psi_n,
\end{equation}
and the quasiperiodic potential operation \( \hat{Q}_n \) introduces a site-dependent phase  \( \epsilon_n \):
\begin{equation}
\hat{Q}_n \psi_n = e^{i \epsilon_n} \psi_n,
\end{equation}
where the quasiperiodic phase parameter is explicitly set to $\epsilon_n = -\alpha n$. To generate an ensemble of trajectories a global offset $\beta$ may be included, $\epsilon_n=-\alpha n+\beta$. Unless stated otherwise we set $\beta=0$.
We choose for \( \alpha \) the Golden mean value $\alpha = \pi \cdot \frac{\sqrt{5} - 1}{2}$, which is irrational, ensuring that the phase \( e^{-i \epsilon_n} \) exhibits quasiperiodic modulations in space.


Application of $\hat{U}_n$ to \(\boldsymbol{\psi}\) yields for the components
$\psi_n(t+1)$ and $\psi_{n+1}(t+1)$
\begin{widetext}
\begin{equation}
\begin{aligned}
\psi_{n}(t+1)
&=
e^{i\epsilon_{n}}
e^{ig|\varphi_n(t)|^2}
\varphi_n(t),\\
\varphi_n(t)
&=
\sin^2\theta\,\psi_{n-2}(t)
-\cos\theta\sin\theta\,\psi_{n-1}(t)
+\cos^2\theta\,\psi_n(t)
+\sin\theta\cos\theta\,\psi_{n+1}(t),\\[2mm]
\psi_{n+1}(t+1)
&=
e^{i\epsilon_{n+1}}
e^{ig|\varphi_{n+1}(t)|^2}
\varphi_{n+1}(t),\\
\varphi_{n+1}(t)
&=
-\sin\theta\cos\theta\,\psi_n(t)
+\cos^2\theta\,\psi_{n+1}(t)
+\sin\theta\cos\theta\,\psi_{n+2}(t)
+\sin^2\theta\,\psi_{n+3}(t).
\end{aligned}
\label{eq1.4}
\end{equation}
\end{widetext}

The map conserves the total squared norm \( \mathcal{A} = \sum_{n=1}^{2N} |\psi_n(t)|^2 \). The local norm density \( |\psi_n|^2 \) is drawn from a Gibbs distribution \( \rho(x) = 2e^{-2x} \) with an average squared-norm density of \( a = \mathcal{A}/(2N) = \frac{1}{2} \) \cite{huang_statistical_1987,rasmussen_statistical_2000}, and the phases are chosen from a uniform distribution on the interval \([0, 2\pi]\). The parameter \(\theta\) controls the coupling strength between neighboring sites.

\section{Localization in Quasiperiodic Unitaries}

In the absence of nonlinearity $g=0$ the linear evolution problem can be mapped onto an eigenvalue problem with eigenvalues densely filling the unit circle and all eigenvectors being exponentially localized. In this linear regime, the time evolution operator 
\(\hat{U} = e^{i\epsilon_n}\hat{C}_{n-1,n}\hat{C}_{n,n+1}\)
acts on a stationary $2N$ dimensional eigenvector \(\phi_m\) as
\begin{equation}
\hat{U}\phi_m = e^{i\omega_m}\phi_m,
\end{equation}
where \(e^{i\omega_m}\) lies on the unit circle and \(\phi_m\) denotes the corresponding eigenstate.

The localization length \(\xi\), which characterizes the spatial decay of wave functions \cite{mallick2023intermediate}, is determined by \(\theta\) and valid for all eigenstates,
\begin{equation}\label{eq1.6}
\xi = \frac{1}{|\ln (|\sin \theta|)|}.
\end{equation}
This implies that smaller \(\theta\) values, \(\theta\simeq 0\), lead to shorter localization lengths, that is, enhanced localization, while larger \(\theta\) values, \(\theta\simeq \pi/2\), result in larger localization lengths, that is, less localized behavior. This is intuitively clear, since in the limit \(\theta=0\) the coupling between the sites \(\psi_i\) vanishes and the modes become localized on a single site, whereas for large values of \(\theta\) the coupling grows and the eigenmodes are spread over increasingly more sites. This is in line with the Anderson localization when keeping the strength of disorder fixed but when varying the strength of coupling (hopping) - for hopping strength going to zero the Anderson eigenmodes will localize on just one lattice site, therefore the localization length is tending to zero in that limit.

Understanding the localization properties of a system requires a quantitative measure for the spatial distribution of the eigenmodes \(\phi_m\). One commonly used metric is the participation number \( P_N \) \cite{skokos_delocalization_2009}, defined as
\begin{equation}
P_N = \frac{1}{\sum_\nu |\phi_{m,\nu}|^4},
\end{equation} 
where $\phi_{m,\nu}$ denotes the $\nu$th component of the eigenvector $\phi_m$. The participation number $P_N$ provides insight into the extent of localization through mainly quantifying its core amplitude distribution: lower values indicate stronger localization, while higher values suggest more extended states. To systematically explore these properties, we compute \( P_N \) across a range of localization lengths \( \xi \) in the absence of nonlinearity.
   
\begin{figure}[!htbp]
    \includegraphics[width=0.5\textwidth]{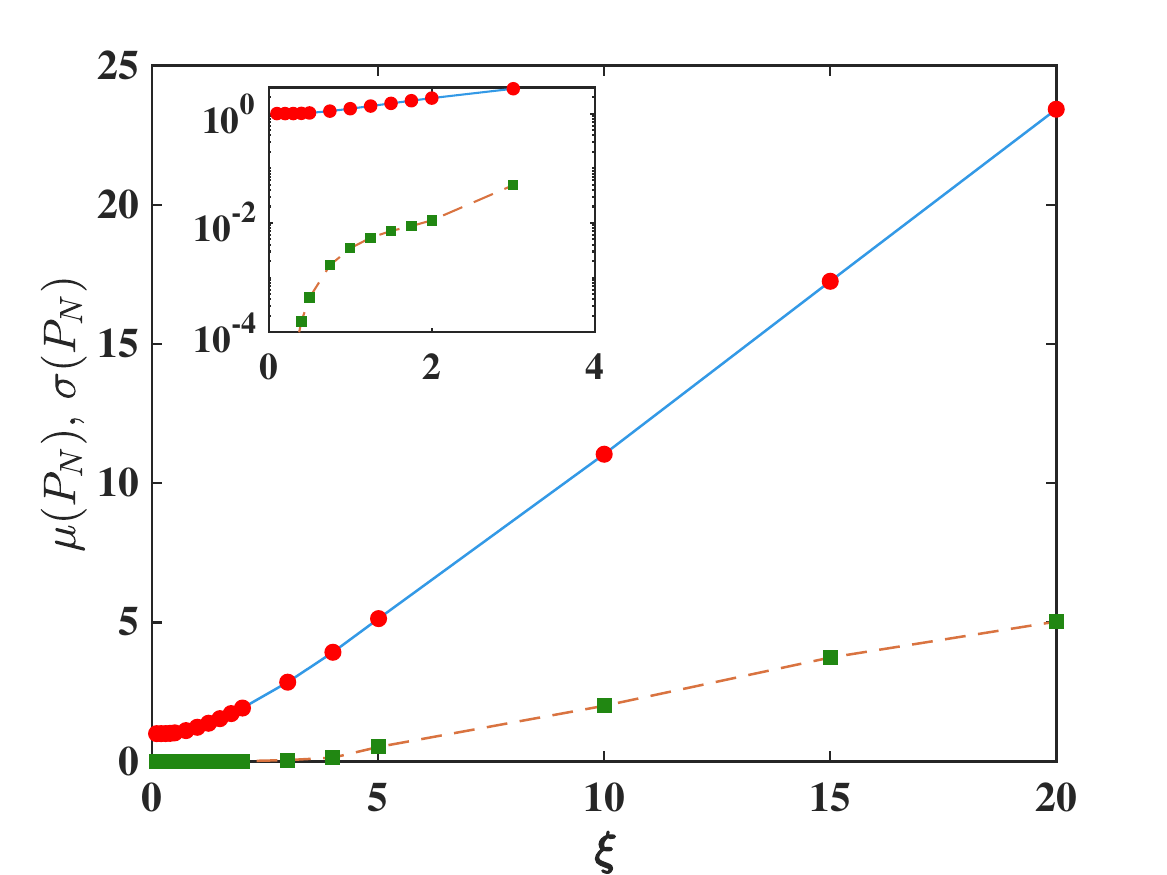}
\caption{
Participation number $P_N$ as a function of the localization length $\xi$ at $g=0$. 
The red circles show the mean value $\mu(P_N)$, and the green squares exhibit the standard deviation $\sigma(P_N)$. 
The inset shows a zoomed-in view of the small-$\xi$ regime, where the $y$ axis is plotted on a logarithmic scale. 
The localization length is sampled at $\xi = 0.1, 0.2, 0.3, 0.4, 0.5, 0.75, 1, 1.25, 1.5, 1.75, 2, 3, 4, 5, 10, 15, 20$.
The system consists of $N = 100$ unit cells.
}
\label{fig:PN_xi}
\end{figure}

Figure~\ref{fig:PN_xi} presents the dependence of the participation number \( P_N \) on the localization length \( \xi \) for the linear case, \( g = 0 \). The red circles and green squares represent the mean participation number \( \mu(P_N) \) and the standard deviation \( \sigma(P_N) \), respectively. Here, averages are taken over the participation numbers of the $2N$ eigenstates. 
As $\xi$ decreases, the mean value $\mu(P_N)$ approaches 1, indicating strong localization of the eigenstates. This behavior is consistent with the theoretical expectation that for sufficiently small $\xi$ eigenstates become exponentially localized, leading to an effective confinement within a single unit cell. The nearly constant value of $\mu(P_N)$ for small $\xi$ further confirms the robustness of quasiperiodic localization in this regime.  

The standard deviation $\sigma(P_N)$ remains at least one order of magnitude smaller than the mean $\mu(P_N)$ for all values of $\xi$, implying that the localization properties are nearly uniform across different eigenstates. This suggests that in the quasiperiodic system, eigenstate localization does not exhibit strong fluctuations, reinforcing the notion that quasiperiodic localization follows an almost deterministic structure rather than a statistical one. 

\subsection{Comparison of quasiperiodic and disorder-induced localization}   

 \begin{figure}[!htbp]
\includegraphics[width=0.5\textwidth]{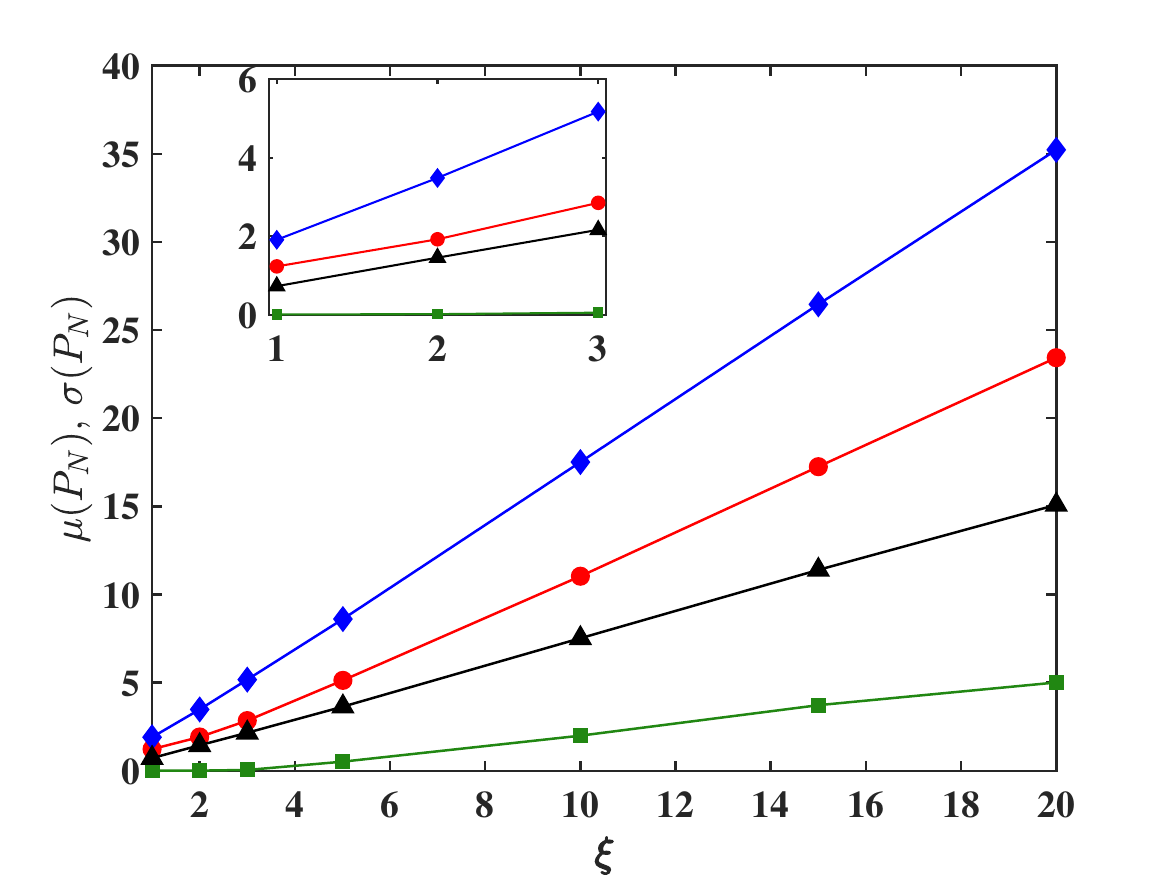}
\caption{Participation number $P_N$ as a function of the localization length $\xi$ for quasiperiodic and disorder systems for the linear case, $g=0$.
The quasiperiodic data are shown as red circles for the mean value $\mu(P_N)$ and green squares for the standard deviation $\sigma(P_N)$, following the notation in Fig.~2. 
The disorder results are indicated by blue diamonds for $\mu(P_N)$ and black triangles for $\sigma(P_N)$. 
The inset displays a magnified view in the small-$\xi$ regime ($\xi = 1, 2, 3$). 
All data are obtained for a system of size $N = 100$.
}
 \label{fig:PN_xi5}
\end{figure}

Let us compare for the linear case $g=0$ the results for the quasiperiodic potential to those for random uncorrelated Anderson disorder, which is obtained by assuming $\epsilon_n$ to be uncorrelated random phases from the entire interval $[ 0,2\pi ]$. The eigenstates are again asymptotically localized exponentially with the localization length dependence on the parameter $\theta$ being exactly the same as in Eq.(\ref{eq1.6}) \cite{vakulchyk_universal_2023}.
To analyze the eigenstate differences between the disordered and quasiperiodic cases, we compute the participation number $P_N$ for ensembles of $N=100$ realizations as function of the localization length $\xi$, as shown in Fig.~\ref{fig:PN_xi5}. The quasiperiodic ensemble is constructed by keeping  $\alpha=\pi(\sqrt{5}-1)/2$ fixed, and changing the global phase, $\epsilon_n=-\alpha n+\beta_m$, where $\beta_m=2\pi m/(2N)$ and $m=1,\ldots,2N$. Furthermore, each realization  of the disorder ensemble is constructed by choosing for each of $2N$ realizations the values of the uncorrelated random phases $\epsilon_n$ independently from the interval $[0,2\pi]$.  All $2N$ eigenstates are included for the computation of $P_N$, so for each value of $\xi$ the average is taken over $(2N)^2$ $P_N$. The mean $\mu(P_N)$ and standard deviation $\sigma(P_N)$ are shown for various values of $\xi$ in Fig.~\ref{fig:PN_xi5}.  

Figure~\ref{fig:PN_xi5} shows that, in the linear regime, the mean participation number $\mu(P_N)$ is consistently higher for the disordered system (blue curve) compared to the quasiperiodic system (red curve) at the same value of $\xi$. This indicates that, on average, eigenstates in the disordered system are more extended than those in the quasiperiodic system. Additionally, the standard deviation $\sigma(P_N)$ (black curve for disorder and green curve for quasiperiodicity) is significantly larger for the disordered case. The same holds for their ratios, $\sigma(P_N)/\mu(P_N)$, implying that the localization properties of eigenstates in the disordered system exhibit larger variability across different realizations.

The results suggest that the quasiperiodic system exhibits stronger localization compared to the disordered system at the same value of $\xi$. The lower mean $P_N$ and smaller standard deviation in the quasiperiodic case indicate that its localization properties are more uniform and deterministic, whereas disorder-induced localization fluctuates due to randomness in the potential. This highlights the fundamental difference between quasiperiodic and disorder-induced localization: quasiperiodic localization is rigid and predictable, while disorder localization exhibits statistical variation.

We therefore expected at a given value of $\xi$ an earlier crossover from LRN to SRN for the quasiperiodic case as compared to the random Anderson one.

\section{Lyapunov spectra}

To probe thermalization slowing down, we analyze the Lyapunov spectrum, which quantifies the exponential divergence of nearby trajectories in phase space.
To compute the Lyapunov exponents, we follow the standard tangent-space approach. The system trajectory $\boldsymbol{\Psi}(t)$ is decomposed into the unperturbed trajectory $\boldsymbol{\psi}(t)$ and a small deviation $\boldsymbol{W}(t)$, such that the trajectory can be expressed as $\boldsymbol{\Psi}(t) = \boldsymbol{\psi}(t) + \boldsymbol{W}(t)$. For convenience, we define the two operators, \(\alpha_n\) and \(\alpha_{n+1}\) [cf. Eq.~(\ref{eq1.4})],
\begin{widetext}
\begin{equation}
\begin{aligned}
\alpha_n X_n
&\coloneqq
\sin^2 \theta X_{n-2}
-\cos \theta \sin \theta X_{n-1}
+\cos^2 \theta X_n
+\sin \theta \cos \theta X_{n+1},
\\
\alpha_{n+1} X_{n+1}
&\coloneqq
-\sin \theta \cos \theta X_n
+\cos^2 \theta X_{n+1}
+\sin \theta \cos \theta X_{n+2}
+\sin^2 \theta X_{n+3}.
\label{operator}
\end{aligned}
\end{equation}
\end{widetext}

By expanding the nonlinear exponential term in the evolution equations for $\Psi_n(t)$ and $\Psi_{n+1}(t)$ and retaining only linear terms in the deviations $W_\nu(t)$ ($\nu = n, n+1$), we obtain the linearized update rule~\cite{malishava_lyapunov_2022}

\begin{widetext}
\begin{equation}
\begin{aligned}
W_\nu(t+1)
&=
e^{i\epsilon_\nu}
e^{ig|\alpha_\nu\psi_\nu(t)|^2}
\Bigl(
\alpha_\nu W_\nu(t)
+
ig\,\Delta_\nu(t)\,
\alpha_\nu\psi_\nu(t)
\Bigr),
\\
\Delta_\nu(t)
&=
\alpha_\nu\psi_\nu(t)\,
\bigl(\alpha_\nu W_\nu(t)\bigr)^*
+
\bigl(\alpha_\nu\psi_\nu(t)\bigr)^*
\alpha_\nu W_\nu(t).
\label{eq3.7}
\end{aligned}
\end{equation}
\end{widetext}

The Lyapunov exponents are defined as \(\Lambda_i = \lim_{t \to \infty} \frac{1}{t} \ln \frac{|W_i(t)|}{|W_i(0)|}\) \cite{benettin1980lyapunov,skokos_lyapunov_2010}, where \(|W_i(t)|\) is the magnitude of the \(i\)-th tangent vector at time \(t\). Due to the symplectic flow properties in tangent space, the Lyapunov spectrum is symmetric, with \(\Lambda_i = -\Lambda_{4N-i+1}\), and two vanishing exponents (\(\Lambda_{2N} = \Lambda_{2N+1} = 0\)) because of norm conservation. The rescaled Lyapunov spectrum \(\bar{\Lambda}\) is then obtained by dividing each Lyapunov exponent \(\Lambda_i\) by the largest Lyapunov exponent \(\Lambda_{\text{max}}=\Lambda_1\), yielding \(\bar{\Lambda}_i = \frac{\Lambda_i}{\Lambda_{\text{max}}}\). The rescaling allows a comparison of the relative magnitudes of the Lyapunov exponents across different parameter regimes, particularly as the system approaches integrable limits. Additionally, the index \(i\) of the Lyapunov exponents is divided by the system size \(2N\) to define the variable \(\rho = i/2N\) with \(\rho \in [0, 1]\), thereby facilitating the analysis of the spectrum in the thermodynamic limit (\(N \to \infty\)), where \(\rho = i/2N\) becomes continuous.

The rescaled KS entropy is defined as the average of all positive Lyapunov exponents in units of the largest one \cite{benettin_kolmogorov_1979}, 
\begin{equation}
\kappa
=
\frac{1}{2N\Lambda_{\mathrm{max}}}
\sum_{\Lambda_i>0}\Lambda_i
=
\int_0^1 \lambda(\rho)\,d\rho ,
\end{equation}
which is a suitable statistical measure for spectral scalings at various limits.
A higher $\kappa$ in proximity to an integrable limit indicates LRN scaling,  while a lower $\kappa$ suggests proximity to SRN scaling.

\section{Scaling results}

When a parameter is tuned such that the system is approaching an integrable limit, the largest Lyapunov exponent and with it the entire spectrum must vanish, i.e. scale down to zero. The scaling details are known to fall into two categories, depending on how the weak nonintegrable perturbation is coupling the otherwise preserved integrals of motion of the limiting integrable case \cite{malishava_lyapunov_2022,malishava_thermalization_2022,lando2023thermalization} . For long range networks (LRN) $\Lambda_1 \rightarrow 0$, $\bar{\lambda}(\rho)$ is stationary and $\kappa$ is constant and finite, resulting in a single parameter scaling \cite{malishava_lyapunov_2022,lando2023thermalization}. For a LRN, no matter how small $\Lambda_1$ gets upon approaching the integrable limit, all other Lyapunov exponents are essentially of the same order of magnitude. For a SRN $\Lambda_1 \rightarrow 0$ and $\bar{\lambda}(\rho) \approx {\rm e}^{-\gamma \rho}$ with a diverging exponent $\gamma \rightarrow \infty$ resulting in a two parameter scaling\cite{malishava_lyapunov_2022,lando2023thermalization}. Contrary to the LRN case, the SRN scaling is therefore characterized not only by a vanishing largest Lyapunov exponent $\Lambda_1$, but in addition by a dramatic exponential diminishing of all other Lyapunov exponents relative to the largest one, i.e. by an exponential slowing down or even an almost complete halt of thermalization. In the SRN case the rescaled KS entropy vanishes upon approaching the integrable limit as $\kappa \approx 1/\gamma \rightarrow 0$.

An ordered linear system will have all actions/eigenmodes being extended over the entire system. Weak nonlinearity will therefore impose an LRN. A disordered linear system whose eigenmodes are localized with a characteristic localization length scale $\xi$ can be expected to show LRN scaling if $\xi \gtrsim N/2$ and SRN scaling if $\xi \ll N$. This prediction was tested with uncorrelated random disorder and Anderson localization in Ref. \cite{zhang2024thermalization} (Fig.5. therein) and confirmed numerically with its onset around $\xi \gtrsim 10$  for a system size $N=200$. There $P_N\approx 18$ as seen from Fig.\ref{fig:PN_xi5}, whereas the quasiperiodic linear case results in a value of $P_N \approx 10$ for the same value of $\xi\simeq 10$. Accordingly, we expect that the transition from LRN to SRN scaling should be even more pronounced and happen at even larger values of $\xi$ as compared to the random Anderson disorder case.

\begin{figure}[!htbp]
    \includegraphics[width=0.45\textwidth]{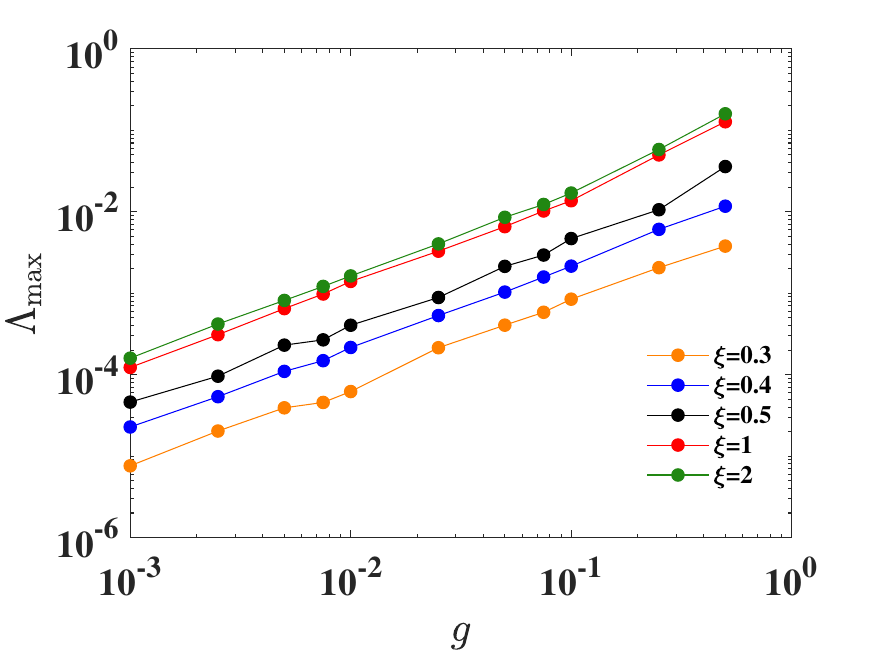}
    \caption{The largest Lyapunov exponent $\Lambda_{\text{max}}$ as a function of the nonlinearity strength $g$ for the quasiperiodic case. Different curves correspond to various localization lengths $\xi$, as indicated in the legend. Data points are sampled for $g = 0.001, 0.0025, 0.005, 0.0075, 0.01, 0.025, 0.05, 0.075, 0.1, 0.25, 0.5$. The scaling behavior of $\Lambda_{\text{max}}$ varies systematically with $\xi$. The system size is $N = 100$, and the evolution time is $10^8$.
}
\label{fig:Lambda_max}
\end{figure}

Figure~\ref{fig:Lambda_max} shows for the quasiperiodic case the dependence of the largest Lyapunov exponent $\Lambda_{\max}$ on the nonlinearity strength $g$ for various localization lengths $\xi$, as indicated in the legend. The data points correspond to different values of $g$, spanning three orders of magnitude. For all values of $\xi$, $\Lambda_{\max}$ decreases as expected monotonically with decreasing $g$, indicating a systematic suppression of chaos as nonlinearity weakens.

 \begin{figure}[!htbp]
    \includegraphics[width=0.45\textwidth]{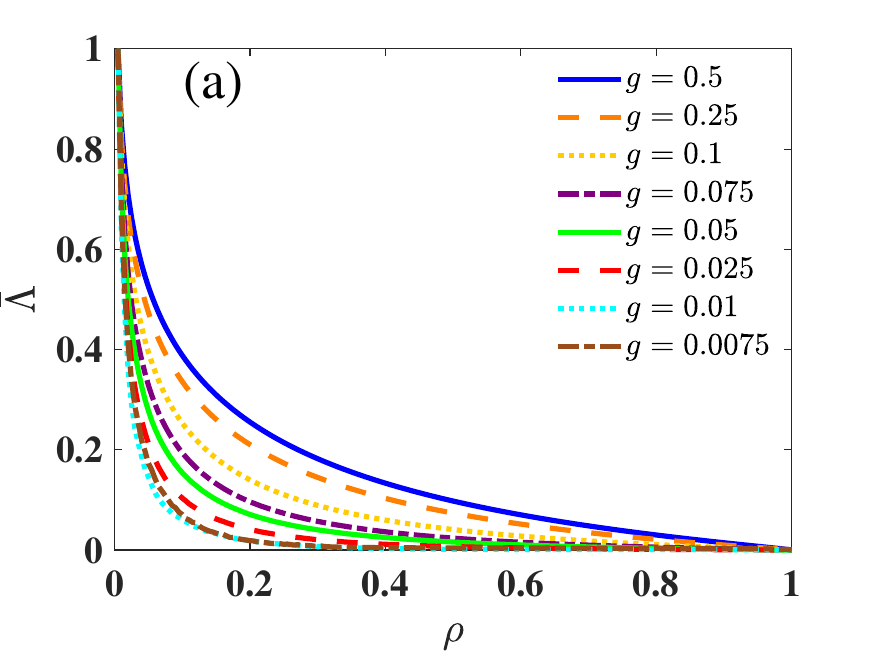}
    \includegraphics[width=0.45\textwidth]{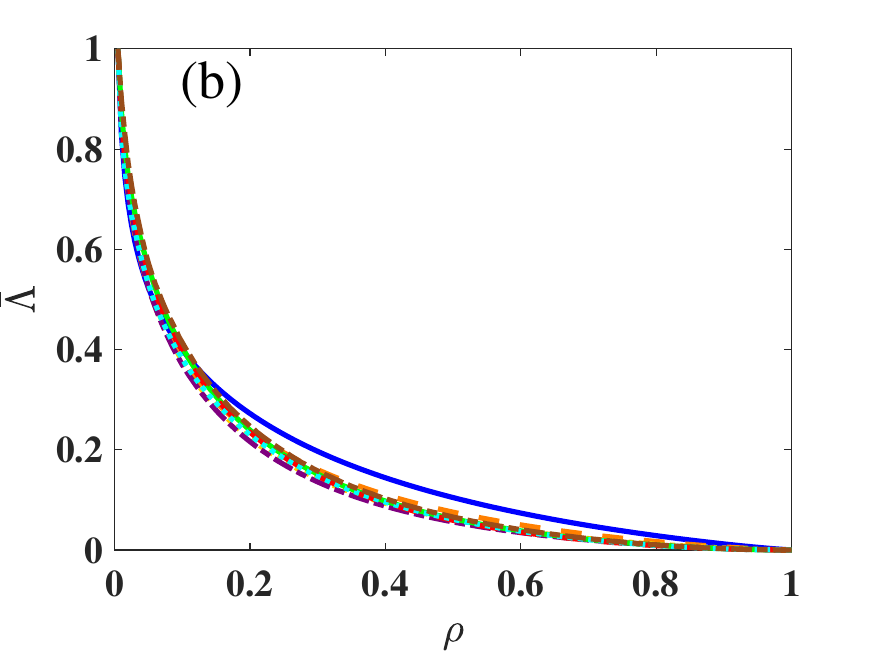}
    \caption{The rescaled Lyapunov spectrum $\bar{\Lambda}$ as a function of the rescaled position $\rho$. Panel (a) corresponds to the disorder case, while panel (b) illustrates the quasiperiodic potential case. Different curves represent varying nonlinearity strengths $g$, ranging from $g = 0.0075$ to $g = 0.5$ (see legend in panel (a)). In both cases, the localization length is fixed at $\xi = 1$, the system size is $N = 100$, and the evolution time is $10^8$. }
    \label{fig:Lambda_rescaled}
\end{figure}

Figure~\ref{fig:Lambda_rescaled} presents the rescaled Lyapunov spectrum $\bar{\Lambda}$ as a function of the rescaled index $\rho$ for a fixed localization length $\xi = 1$. Panel (a) corresponds to the disordered case, while panel (b) illustrates the quasiperiodic potential case. Different curves represent varying nonlinearity strengths $g$, ranging from $g = 0.0075$ to $g = 0.5$.  Surprisingly we observe a much stronger tendency towards SRN for the disordered case, while the quasiperiodic case appears to show LRN scaling.
While the disordered case results in spectra which bend faster and faster down to that for the smallest computed values of $g$ [panel (a)], the quasiperiodic system [panel (b)] results in rescaled Lyapunov spectra which collapse onto a stationary curve for all but the largest values of $g$. 
This is contrary to the above expectation that the quasiperiodic case should show stronger SRN scaling for the same value of $\xi \ll N$. 

\begin{figure}[!htbp]
\includegraphics[width=0.45\textwidth]{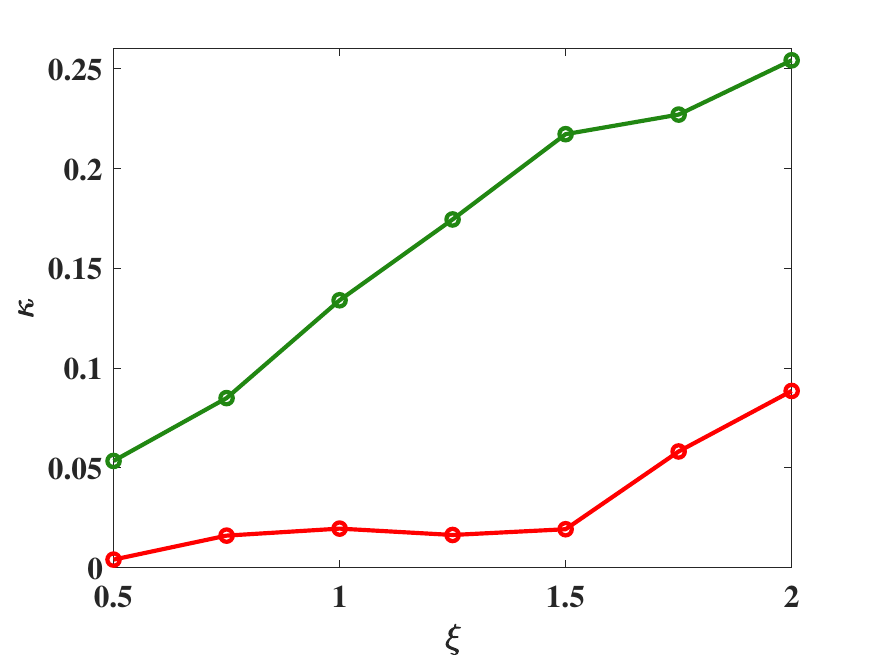}
    \caption{The rescaled Kolmogorov-Sinai (KS) entropy $\kappa$ for $g=0.0075$ 
    as a function of the localization length $\xi$. The green curve represents the quasiperiodic case, while the red curve corresponds to the disordered case. Data points are sampled at $\xi = 0.5, 0.75, 1, 1.25, 1.5, 1.75, 2$.
    }
\label{fig:KS_entropy}
\end{figure}

In order to quantify the surprising scaling difference between the disordered and quasiperiodic cases, 
we compute the rescaled Kolmogorov-Sinai (KS) entropy $\kappa$ as a function of the localization length $\xi$ in proximity to the integrable limit, as shown in Fig.~\ref{fig:KS_entropy}, 
and compare the KS entropy for the quasiperiodic (green curve) and disordered (red curve) cases. Here, $\kappa$ is obtained from the smallest value of the integrability breaking nonlinearity $g$. Clear cut LRN values would be $\kappa \gtrsim 0.1$.
The disorder case shows a decrease of $\kappa$ with $\xi$ from LRN values at $\xi \approx 2$ down to SRN values of about $\kappa \approx 0.02$ and less, once the localization length decreases beyond $\xi \leq 1.5$. 
The quasiperiodic case however shows LRN values of $\kappa$ down to $\xi \approx 0.8$. Only for way smaller values $\xi \leq 0.5$ is the quasiperiodic case indicating a clear transiting into SRN scaling. A rough linear extrapolation of the green curve in Fig.\ref{fig:KS_entropy} results in a value of about $\xi=0.1$ for the transition, which is one order of magnitude smaller than that for the disorder case, $\xi=1.5$.

\section{conclusion}

We aimed at a clear observation of LRN to SRN scaling transitions for thermalization of quasiperiodic weakly nonlinear unitary maps. In particular we expected to observe this transition at larger values of the localization length as compared to previously observed transitions for random disorder instead of quasiperiodicity. To our surprise we report that the quasiperiodic case transits, but does so for much smaller values of the localization length, approximately one order of magnitude smaller than for the disorder case. 
A possible reason could be the presence of long-range correlations in the linear eigenfunctions of the quasiperiodic case, which are present in the chaotic thermalization dynamics and delay the transition. Future comparative studies could examine the time-continuous case of Gross-Pitaevski lattices with disorder, corresponding to Anderson localization for the linear case, and quasiperiodic potentials, corresponding to Aubry-Andre localization for the linear case, in order to possibly confirm or modify our above insights. Insights from the purely linear cases are provided in Appendix A.

\section*{Acknowledgement}
This research was supported by the Institute for Basic Science
through Project Code (No. IBS-R024-D1 and IBS-R041-D1-2026-a00). X.Z. acknowledges the financial support from the NSF of China (Grant No. 12247101), the 111 Project (Grant No. B20063), and the China Scholarship Council (Grant No. CSC-202306180087). BD was partially funded by the Deutsche Forschungsgemeinschaft (DFG, German Research Foundation) Project No. 290128388. We gratefully acknowledge helpful discussions with Gabriel Lando.

\appendix 
\section{Comparison of Quasiperiodic and Disorder-Induced Localization in a Tight-Binding Model} 

To complement the analysis of localization in nonlinear unitary maps, we also examine a standard one-dimensional tight-binding model in the linear regime, which provides a well-understood framework to compare quasiperiodic and disorder-induced localization. The model is governed by the equation
\begin{equation}
E \psi_\ell = \epsilon_\ell \psi_\ell - t (\psi_{\ell+1} + \psi_{\ell-1}),
\end{equation}
where $\psi_\ell$ is the wave-function amplitude at site $\ell$, and $t$ is the hopping amplitude, which is set to $t=1$ here. Periodic boundary conditions are applied, i.e., $\psi_{\ell+1} = \psi_1$. The potential $\epsilon_\ell$ is specified in two distinct forms. In the disordered case, $\epsilon_\ell$ is drawn from a uniform distribution in the interval $\left[-W/2, W/2\right]$, and the corresponding localization length is approximately given by $\xi(W) \approx 100/W^2$ for $W \lesssim 4$ \cite{anderson_absence_1958,vakulchyk_anderson_2017}. In the quasiperiodic case, the potential takes the form $\epsilon_\ell = \lambda \cos(\beta + 2\pi \alpha \ell)$, where $\alpha = (\sqrt{5} - 1)/2$ is an irrational number and $\beta$ is a global phase. For $\lambda > 2$, the localization length follows $\xi = 1/\ln(\lambda/2) $\cite{aubry1980analyticity}.

 \begin{figure}[!htbp]
    \includegraphics[width=0.45\textwidth]{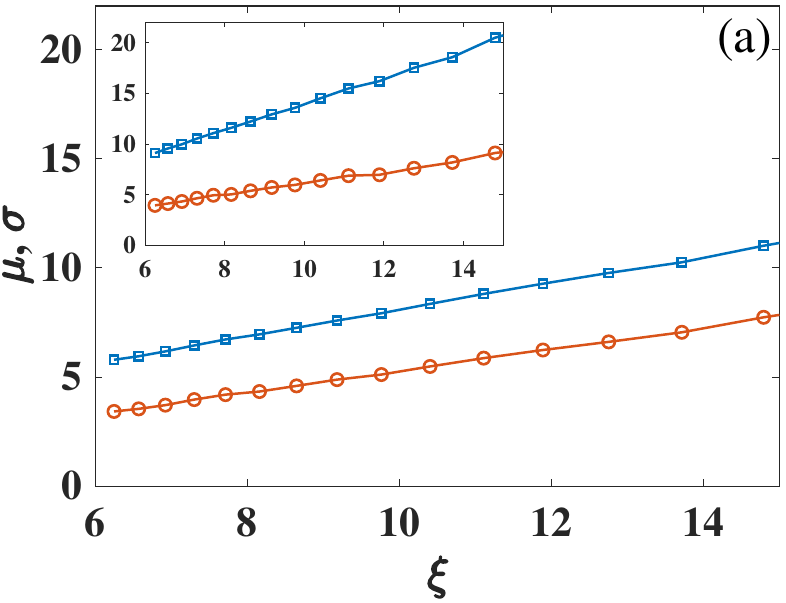}
     \includegraphics[width=0.45\textwidth]{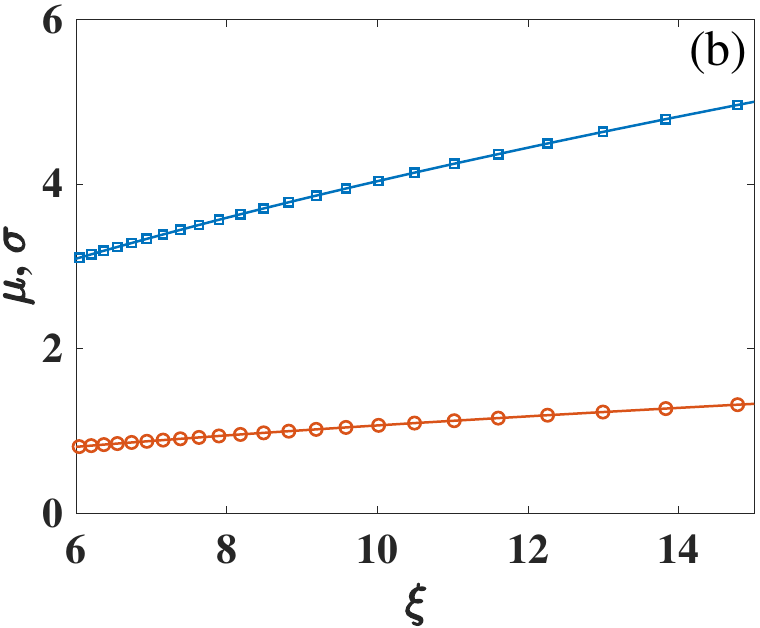}
    \caption{Average participation number $\mu$ (blue squares) and standard deviation $\sigma$ (orange circles) as functions of localization length $\xi$ in a one-dimensional tight-binding model. (a) Disordered case. The main panel shows data for eigenstates with $|E|<1$, while the inset shows results for $|E|<0.1$. (b) Quasiperiodic potential case. 
    }
    \label{P10}
\end{figure}

We compute the participation number $PN$ as a function of the localization length $\xi$ in both the disordered and quasiperiodic cases. For the disordered system, results are averaged over multiple realizations of disorder, while for the quasiperiodic system, averages are taken over different global phases $\beta$. As shown in Fig.~\ref{P10}, the quasiperiodic case exhibits lower mean values of $PN$ and smaller standard deviations than the disordered case, particularly at small $\xi$. These results indicate stronger and more uniform localization in the quasiperiodic system, consistent with its deterministic modulation. In contrast, the disordered system shows larger variability, reflecting its statistical nature. This comparison reinforces the conclusions drawn from the linear unitary map model in the main text, namely, that quasiperiodic localization is more structurally rigid and less sensitive to fluctuations than disorder-induced localization.

\newpage

\bibliography{sergejflach}

\end{document}